\documentclass[runningheads]{llncs}
\usepackage{graphicx}

\usepackage{tikz}
\usepackage{multirow}
\usepackage{comment}
\usepackage{amsmath,amssymb} 
\usepackage{color}
\usepackage{xspace}
\usepackage{subcaption}
\usepackage{array}
\usepackage{arydshln}
\makeatletter
\DeclareRobustCommand\onedot{\futurelet\@let@token\@onedot}
\def\@onedot{\ifx\@let@token.\else.\null\fi\xspace}

\def\eg{\emph{e.g}\onedot} 
\def\ie{\emph{i.e}\onedot} 
 
\def\etc{\emph{etc}\onedot} 
 
\def\etal{\emph{et al}\onedot}
\makeatother

\definecolor{lilicolor}{rgb}{0.1,0.9,0.1}
\newcommand{\lili}[1]{{\textcolor{lilicolor}{[LT #1]}}}

\definecolor{seancolor}{rgb}{0.1,0.1,0.9}

\definecolor{francolor}{rgb}{0.858,0.188,0.478}

\definecolor{ruibocolor}{rgb}{0.358,0.588,0.778}
\newcommand{\ruibo}[1]{{\textcolor{ruibocolor}{[RS #1]}}}

\definecolor{rohancolor}{rgb}{0.9,0.1,0.1}
\newcommand{\rohan}[1]{{\textcolor{rohancolor}{[RS #1]}}}

\definecolor{todocolor}{rgb}{0.9,0.1,0.1}

\renewcommand{\ruibo}[1]{#1}

\renewcommand{\rohan}[1]{#1}
\renewcommand{\lili}[1]{#1}

\usepackage[accsupp]{axessibility}  

\begin{document}
\pagestyle{headings}
\mainmatter
\def\ECCVSubNumber{4164}  

\title{CV4Code: Sourcecode Understanding via Visual Code Representations} 

\titlerunning{CV4Code: Visual Sourcecode Representations}
%
\author{Ruibo Shi\and
Lili Tao\and
Rohan Saphal\and
Fran Silavong\and
Sean J. Moran}
\authorrunning{R. Shi \and L. Tao et al.}
%
\institute{CTO, JP Morgan Chase}
\maketitle

\begin{abstract}
We present CV4Code, a compact and effective \emph{computer vision} method for sourcecode understanding. Our method leverages the contextual and the structural information available from the code snippet by treating each snippet as a two-dimensional image, which naturally encodes the context and retains the underlying structural information through an explicit spatial representation. To codify snippets as images, we propose an ASCII codepoint-based image representation that facilitates fast generation of sourcecode images and eliminates redundancy in the encoding that would arise from an RGB pixel representation. Furthermore, as sourcecode is treated as images, neither lexical analysis (tokenisation) nor syntax tree parsing is required, which makes the proposed method agnostic to any particular programming language and lightweight from the application pipeline point of view. CV4Code can even featurise syntactically incorrect code which is not possible from methods that depend on the Abstract Syntax Tree (AST). We demonstrate the effectiveness of CV4Code by learning Convolutional and Transformer networks to predict the functional task, \ie~ the problem it solves, of the source code directly from its two-dimensional representation, and using an embedding from its latent space to derive a similarity score of two code snippets in a retrieval setup. Experimental results show that our approach achieves state-of-the-art performance in comparison to other methods with the same task and data configurations. For the first time we show the benefits of treating sourcecode understanding as a form of image processing task.
\end{abstract}
\section{Introduction}
\label{sec:intro}

Machine Learning on Sourcecode (MLOnCode) promises to redefine how software is delivered through intelligent augmentation of the software development lifecycle (SDLC). Automation of routine tasks with software makes our lives more comfortable and efficient. For example, software drives the global economy and transfer of value worldwide making the purchase of goods and the management of finances a seamless experience. Furthermore, at the touch of a few buttons on our smartphone we can communicate to friends and family worldwide. Reliable software can also change healthcare outcomes by making it easier for doctors to diagnose disease.
A key to accomplishing these feats of automation and being prepared for future complex use-cases is accelerating the software development process without sacrificing software quality, robustness and time-to-market. Augmenting the SDLC with machine learning holds this promise, in which the developer's capabilities are magnified through predictive analytics  driven by the vast quantities of exhaust data naturally produced by the SDLC. Machine learning can potentially enhance every stage of the SDLC including requirements gathering, build and test and deployment. For example, AI-driven code auto-completion and enhanced code search are near-term possibilities for enhancing developer productivity with startups and established companies alike productionising such capabilities for mass consumption\footnote{\url{https://copilot.github.com/}}.
\begin{figure}[t]
    \centering
    \includegraphics[width=0.6\textwidth]{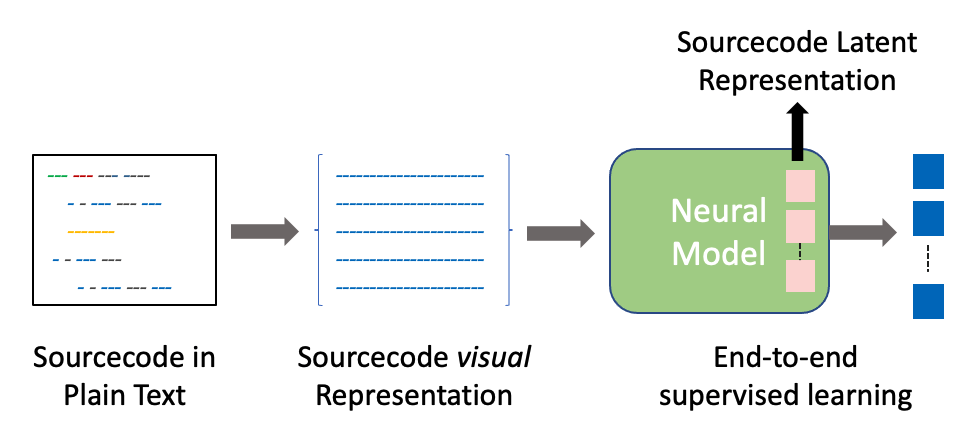}
    \caption{The proposed CV4Code code understanding pipeline. }
    \label{fig:workflow}
\end{figure}

The academic field of MLOnCode explores the application of machine learning techniques for mining the massive amount of sourcecode and associated metadata available in public and private repositories~\cite{allamanis2018survey}. Indicative tasks in this field include code search using natural language keywords~\cite{Shuhan20} and sourcecode~\cite{luan2019} as queries, automated bug finding~\cite{Pradel2018}, vulnerability detection~\cite{Russell18}, design pattern detection~\cite{Zanoni15}, program repair~\cite{Chen21} and code auto-completion~\cite{Svyatkovskiy21}. Core to the field of MLOnCode is the learning of expressive sourcecode latent feature representations (``code vectors'') that capture semantics of programs and can be flexibly used in generic machine learning classifiers to support a myriad of downstream tasks, such as code search and repository annotation with semantic keywords. Code is unique from natural language in many respects, for example more distantly spaced tokens may be highly related (\eg opening and closing brackets) and code that looks very similar can have very different behaviour (and vice-versa). Capturing these subtleties and intricacies of sourcecode to learn program semantics requires methods that can understand the underlying context (sequence of tokens) and structure (as presented by syntax parse tree) of the language. Prior methods for sourcecode feature extraction can be differentiated to the extent on which they capture context, structure or both when learning code vectors. For example, early methods treat sourcecode either as a set of independent tokens~\cite{Allamanis13}, a sequence of tokens (processed sequentially by an RNN or CNN) or generate an Abstract Syntax Tree (AST) from the code snippet before linearising the tree into vector form, thereby capturing local structure in the snippet~\cite{luan2019}. Drawing a parallel between the word2vec model in natural language processing (NLP), Alon~\etal~\cite{Uri19} propose a code2vec alternative that uses the proxy task of method name prediction based on paths in the AST tree to learn expressive code vectors from a shallow neural network architecture. More recent research has explored transformer architectures to learn effective code vectors~\cite{zuegner_code_transformer_2021} that capture structure and context. In contrast to prior research, we represent sourcecode in a visual way as a set of \emph{images} that explicitly, through the 2-dimensional spatial representation, present both the code structure and context directly to the learning algorithm. We adapt well-known image processing techniques to process these sourcecode images. We argue that, with no assumption of naturalness in programming language~\cite{allamanis2018survey}, treating sourcecode understanding as a computer vision problem can not only produce more effective code vectors, but can also address key limitations of existing methods such as their inability to featurise partial code snippets and syntactically incorrect code.

In more detail we introduce a series of vision models, including Residual Convolutional Neural Networks (CNNs) \cite{he2016residual} and Vision Transformers (ViT) variants ~\cite{alex2021vit,Lee2021VisionTF,Hassani2021EscapingTB}, adapted for sourcecode understanding (Figure~\ref{fig:workflow}). We contribute to the sparse amount of prior research that draws a parallel between successful image understanding models in the field of Computer Vision and their application to representation learning for sourcecode~\cite{bilgin2021code2image,rabin2021encoding,dey2019socodecnn}. Different to this closely related prior research, CV4Code does not require any language-specific pre-processing (\eg~extraction of syntax parse trees). To represent sourcecode in a visual form, we propose a novel compact encoding of sourcecode as a two dimensional spatial grid of numeric values that represent the characters in the code by their ASCII codepoints. This representation is advantageous over a standard RGB pixel representation of the code for two key reasons:~1)~\emph{elimination of redundancy;} for a pixel representation many pixels would be dedicated to encoding a single character; and 2)~\emph{fast feature generation:} we find it multiple order of magnitude slower and less scalable to render a pixel representation of code (sub-second) compared to the proposed code representation (sub-millisecond). The computational speed-up enables real-time applications over large codebases, such as code search directly within the Integrated Development Environment (IDE). The proposed image encoding for sourcecode is therefore practical for real-world machine learning pipelines. The CV4Code architecture is designed for learning representations effectively from the ASCII-based codepoint encoded sourcecode images. CV4Code ingests the image representation of code and encodes each pixel as either a one-hot or learnable embedding. The encoded input is subsequently processed by a neural network that learns features expressive for the predictive task~\eg the language of code, the task being solved by the code~\etc. And the learned latent embedding from CV4Code can be used as code embeddings for other MLOnCode tasks, similar to VGG features~\cite{Simonyan15} that have been shown to be a powerful and flexible embedding of images for many computer vision tasks.

Our contributions in this paper can be summarised as:
\begin{itemize}
    \item \textbf{CV4Code Deep Neural Models:} We introduce and compare modern deep vision models adapted for language-agnostic sourcecode understanding that learns from a novel ASCII codepoint representation of sourcecode. We report state-of-the-art performance on a public benchmark dataset compared to competitive baselines. Compared to a NLP-inspired Transformer strong baseline, we achieve $4.06\%$ absolute gain in top-1 accuracy on a language-agnostic problem classification task, and 0.011 gain in mAP@R on a similarity based sourcecode retrieval task.
    \item \textbf{ASCII Sourcecode Encoding:} We introduce an ASCII codepoint image representation for sourcecode that efficiently (low redundancy, fast generation) encodes snippets capturing both code structure and context in a single representation.
\end{itemize}
\section{Related work}
\label{sec:related_work}
\lili{Machine learning for sourcecode analysis aims at learning semantically meaningful representation of the code and then apply the embeddings on downstream tasks, such as code quality detection, code summarisation, defect prediction and code duplication detection~\cite{sharma2021survey}. We include the research that are most relevant to our contribution.}

\noindent \textbf{AST based representation. }
\lili{Machine learning based intelligent code analysis relies on extracting representative features from sourcecode. The majority of studies leverage structured graphical models for sourcecode, through parse trees. AST carries rich semantic and syntactic information and provides a unique representation of a sourcecode snippet in a given language and grammar ~\cite{alon2018codeseq,zhang2019novel,zugner2021language}. The paper ~\cite{zugner2021language} learned jointly from the AST and the sourcecode of programs while relying on language-agnostic features, and performed on code summarisation task on five programming languages. Although the model does not rely on language-specific features, parsing sourcecode to a tree structure is language dependent. }

\noindent \textbf{NLP based techniques. } 
\lili{A sourcecode sample can be treated as a piece of text. A code snippet can be represented by a vector of frequencies of token occurrences, similar to the bag of word model. The frequently used tokens include regex, keywords and operators ~\cite{ochodek2020recognizing,puri2021codenet}. Considering the lack of sequential information retained in the bag of tokens method, a sequence of token method uses the same set of tokens but keeps the order information to form a sequence~\cite{puri2021codenet}. Such a token embedding layer is then input into a CNN based model. }

\noindent \textbf{Computer vision for code. }
\lili{Despite the fact that more effort has been made on automated sourcecode analysis using machine learning, representing and processing sourcecode in the form of image data is still an under-explored area in the field. Dey~\etal~\cite{dey2019socodecnn} automatically convert program sourcecode to visual images via an intermediate representation of code created by the LLVM compiler. The ASCII value of the remaining characters are treated as a pixel value in a predefined empty image canvas. Bilgin~\etal~\cite{bilgin2021code2image} use a coloured image of syntax trees for representing code, where the tokens are plotted in a rectangular shape and are completed with specific colours to indicate the type and content of the token. The most recent work by Rabin~\etal~\cite{rabin2021encoding} on Java code analysis transforms the original code by removing comments and empty lines using a JavaParser tool, and then redacts snapshots of input by replacing any alphanumeric characters in the reformatted input programs with a single letter ‘x’ to emphasise the structure of code snippets, rather than their content (\eg~the specific naming of variables). While the aforementioned research exploited image-based representation for sourcecode, they all require a parser for specific programming language, which cannot process the sentences with incorrect syntax. }

\section{CV4Code }
\label{sec3}
\ruibo{We propose CV4Code, an end-to-end learning framework for sourcecode understanding by treating sourcecode snippets as \emph{images}~\ie.~2-dimensional matrix. }

\subsection{Sourcecode Representation} \label{ssec:sourcecode_representation}
\begin{figure}[h]
    \centering
    \includegraphics[width=0.6\textwidth]{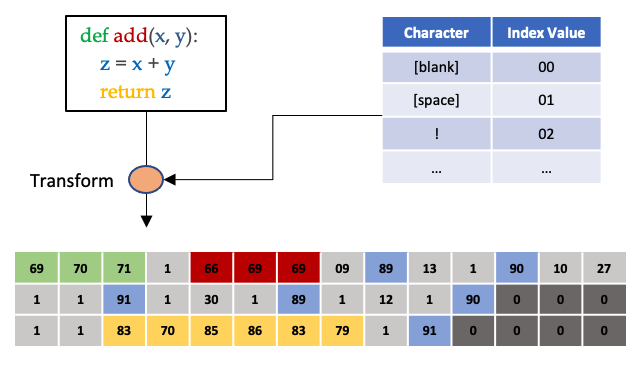}
    \caption{Example of 2D code representation generation}
    \label{fig:code_generation_example}
\end{figure}

\begin{figure*}[t]
\begin{subfigure}{\textwidth}
    \centering
    \includegraphics[width=.9\linewidth]{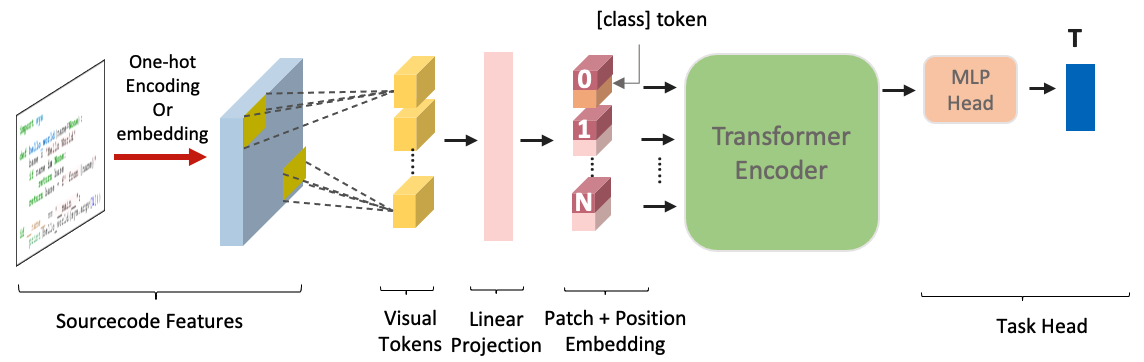}
    \caption{CV4Code ViT model overview. }
    \label{fig:model_overview_vit}
\end{subfigure}
\begin{subfigure}{\textwidth}
    \centering
    \includegraphics[width=.9\linewidth]{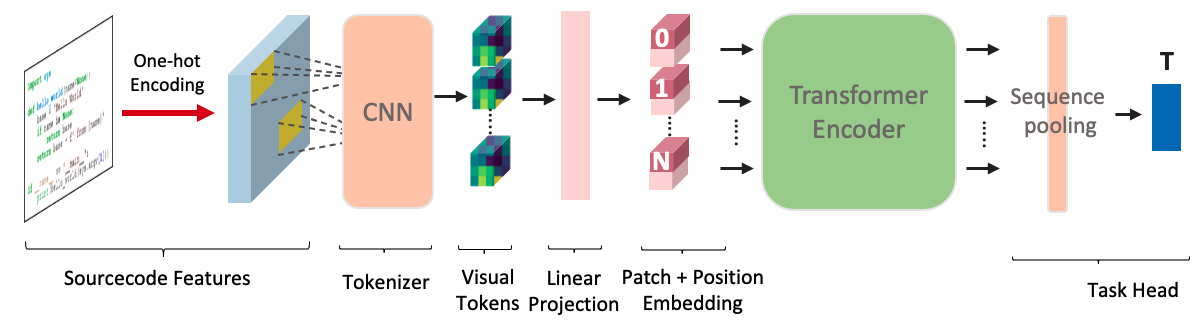}
    \caption{CV4Code CCT model overview. }
    \label{fig:model_overview_cct}
\end{subfigure}
\caption{CV4Code transformer model variants. }
\label{fig:model_overview}
\end{figure*}

\ruibo{
While the sourcecode of most modern programming languages can be written in plain text from an extensive character set, only a small set of tokens and their composing characters have syntactic and semantic roles. In CV4Code, code snippets are transformed into 2-dimensional (matrix) representation by mapping each printable ASCII character to their unique index values and padding the special \emph{[blank]} token wherever necessary to retain the rectangular shape of the output. The set of valid printable ASCII characters together with the special padding token $\mathbb{V}_{c}$, $|\mathbb{V}_{c}|=96$, consists of the following:
\newline
\begin{center}
  \begin{tabular}{l}
    \texttt{abcdefghijklmnopqrstuvwxyz} \\
    \texttt{ABCDEFGHIJKLMNOPQRSTUVWXYZ} \\
    \texttt{0123456789} \\
    \texttt{!"\#\$\%\&'()*+,-./:;<=>?@[\char`\\]$^\wedge$\char`_\char`\`\char`\{\char`\}\char`|$\sim$ } \\
    \texttt{\emph{[space][blank]}} \\
\end{tabular}
\end{center}

Figure \ref{fig:code_generation_example} shows an example of the code representation generation process. Specifically, for a code snippet spanning $L$ lines each with $C_{l}$, $l\in{0, ..., L-1}$ characters, the transformation is done in three steps: 
\begin{enumerate}
    \item Remove characters not within the valid set, output has $\hat{L}$ lines each with $\hat{C}_{l}$, $l\in{0, ..., \hat{L}-1}$ characters;
    \item Map each input character $v_{k} \in \mathbb{V}_{c}$ to its index value $k$;
    \item Pad each line to $M=\max_{l=0}^{\hat{L}-1}\hat{C}_l$ long with the index value of \emph{[blank]}, generate the output 2-dimensional code matrix $X\in\mathbb{R}^{L\times M}$.
\end{enumerate}
}
\ruibo{Compared to human-readable images of sourcecode, ~\eg~screenshots of code snippets, which usually are sparse in semantics and require multiple pixels to represent a single character, the proposed sourcecode representation is compact and do not introduce any unnecessary information other than the blank padding that is required to keep the spatial relations.}

\ruibo{As the code image, \ie~the compact 2-dimensional sourcecode representation, encodes the character index values which do not form a numerically continuous space, unless otherwise specified, one-hot encoding is used in this work to transform each \emph{pixel} in the code image to a vector of fixed dimension equal to the size of the set of valid characters, i.e. $\mathbf{X}\in\mathbb{R}^{L\times M} \rightarrow \mathbf{\hat{X}}\in\mathbb{R}^{L\times M\times |\mathbb{V}_{c}|}$. }

\subsection{Model}
\ruibo{We apply and adapt state-of-the-art vision models, including ResNet~\cite{he2016residual}, ViT~\cite{alex2021vit}, ViT for small-size datasets (ViT-fsd)~\cite{Lee2021VisionTF} and Compact Convolution Transformer (CCT)~\cite{Hassani2021EscapingTB}, on the proposed sourcecode representation. And we show the effectiveness of the proposed method through experiments on a supervised multi-class classification task. Figure \ref{fig:model_overview} shows an overview of the CV4Code transformer model variants. While Figure \ref{fig:model_overview_vit} shows a general architecture of CV4Code-ViT model, differences exist in ViT, ViT-fsd which we briefly describe below. 
\newline\textbf{ViT.} We follow \cite{alex2021vit} and split images into non-overlapping fixed-size patches and   prepend a learnable \emph{[class]} embedding whose state at the ViT output serves as the sourcecode representation.  This sourcecode representation is then passed to a single-layer MLP head for the classification task. 
\newline\textbf{ViT-fsd.} While the same setup as ViT is used, we apply shifted patch tokenization and Locality Self-Attention proposed in \cite{Lee2021VisionTF}. In addition, as the tokenization process creates 4 extra shifted images leading to a largely increased input dimensionality after concatenation, to control the number of parameters in the linear projection layer, instead of one-hot encoding,~\ie $\mathbf{\hat{X}}\in\mathbb{R}^{L\times M\times |\mathbb{V}_{c}|}$, an 32-dimensional learnable embedding is used such that $\mathbf{\hat{X}}\in\mathbb{R}^{L\times M\times 32}$.
\newline\textbf{CCT.} Shown in Figure \ref{fig:model_overview_cct}. To leverage CNN's inductive bias,~\eg locality, a variant of CCT is proposed in this work. Similar to \cite{Hassani2021EscapingTB} we use convolutional layers to create soft visual tokens and replace the use of \emph{[class]} embedding with sequence pooling to generate sourcecode representation at the output. Furthermore, as the soft tokenization process does not require fixed-size input, similar to NLP applications of transformers where the input sequence is of variable length, for smaller sourcecode image we append learnable \emph{[pad]} embeddings to the generated visual token sequence,~\ie at the output of the CNN, to form equal-length input to the transformer encoder.
}
\subsection{Implementation Details} \label{sec:implementation_details}

\noindent\textbf{Variable code snippet size.} \ruibo{It is expected that the sourcecode 2-dimensional representation will vary in size. To address this, using the \emph{[blank]} token, we batch up sourcecode snippets of different sizes with \emph{interleaved} padding vertically and constant padding horizontally. While constant padding appends constant values from the end of an array, \emph{interleaved} padding avoids leaving large continuous blank region by inserting \emph{[blank]} tokens between original input code lines. In contrast, if an image exceeds the maximum size limit, we crop and keep the top left corner of the code image, following the raster order to retain most information. For instance, given input of size $L\times M$, $\mathbf{X}_{i} = [\mathbf{x}_{0}, \mathbf{x}_{1}, ..., \mathbf{x}_{L-1}]$ where $\mathbf{x}_l$ for $l = 0, ..., L-1$ each is a row vector of length $M$, then the cropped output of size $\hat{L} \times \hat{M}$ is $\mathbf{\hat{X}_{i}} = [\mathbf{\hat{x}}_{0}, \mathbf{\hat{x}}_{1}, ..., \mathbf{\hat{x}}_{\hat{L}}]$ where $\mathbf{\hat{x}_{l}}$ = \{$x_{l, m}\}_{m=0}^{\hat{M}}$ for $l = 0, ..., \hat{L}-1$. In addition, we pad or limit the input images to be the same size of $96\times96$ for ResNet, ViT and ViT-fsd, while in CCT, along with an global minimum of $12\times12$ and maximum of $96\times 96$ on all batches, we dynamically limit the maximum image size per minibatch at training time,~\ie instead of setting a constant maximum limit in all minibatches, it is configured to the 95th-percentile, if smaller than the global maximum, of the width and height independently of the images within the minibatch. }

\noindent\textbf{Training.} \ruibo{We use AdamW \cite{Loshchilov2019DecoupledWD} optimiser with an learning rate of $10^{-3}$ and weight decay is set to 0.0001. We also use a 5-epoch warm-up along with a Cosine Learning Rate Annealing. Unless otherwise specified, all models are trained for 100 epochs and the model with the highest validation accuracy is selected to report results on the test set. Our model is implemented in PyTorch and trained on 1x NVIDIA V100 GPU with a batch size of 256.}

\section{Experimental Evaluation}\label{sec:experiments}
\label{sec:results}

 \rohan{In order to benchmark the performance and capabilities of our proposed framework, we conduct experiments on a real-world sourcecode dataset and compare against a set of competitive baselines, including character-, token- and AST-based representations respectively. Furthermore, test results are reported on Code Classification and Code Similarity tasks. The goal of these experiments is to answer the following research questions :
 \begin{itemize}
     \item RQ1: How well does CV4Code perform on the tasks in comparison to baselines using alternative forms of source code representation?
     \item RQ2: How scalable and flexible is CV4Code to baselines using alternative forms of source code representation?
     \item RQ3: How useful are the latent features learnt by CV4Code for alternate downstream tasks?
 \end{itemize}}

\subsection{Setup}
\begin{table}[t]
    \centering
    \begin{tabular}{c|c|c|c}
    \hline
        \multirow{2}{*}{\textbf{Dataset}} & \multicolumn{3}{|c}{\textbf{Summary}} \\
        & \#problems & \#samples & \#languages \\
        \hline\hline
        CodeNetBench-Train & 237 & 171300 & 3 \\
        CodeNetBench-Validation & 237 & 21000 & 3 \\
        CodeNetBench-Test & 237 & 21000 & 3 \\
    \hline
    \end{tabular}
    \caption{CodeNetBench data summary. Balanced distribution among C++, Python and Java.}
    \label{tab:classification_dataset}
\end{table}
\begin{table}[t]
    \centering
    \begin{tabular}{c|c|c|c}
    \hline
        \multirow{2}{*}{\textbf{Dataset}} & \multicolumn{3}{|c}{\textbf{Summary}} \\
        & \#problems & \#samples & \#languages \\
        \hline\hline
        CodeNetBench-Sim & 100 & 2000 & 2  \\
    \hline
    \end{tabular}
    \caption{Similarity evaluation datasets.}
    \label{tab:similarity_dataset}
\end{table}

\noindent\textbf{Dataset. }
\rohan{We use CodeNet \cite{puri2021codenet}, a high-quality real-world dataset with code samples scraped from online coding platforms. The dataset provides code samples submitted by students and developers from across the world, resulting in a sundry pool of source code. The dataset is categorised based on the problems presented in the platform and for each such problem it provides the submissions in multiple programming languages. Our choice of CodeNet stems from the fact that we aim to benchmark the sourcecode understanding capability of CV4Code against baseline models in both language-agnostic and language-specific setups, and high-quality CodeNet benchmark set supporting 3 popular languages, including C++, Java and Python, makes it the ideal choice. }
\ruibo{We use the curated benchmark set of CodeNet \cite{puri2021codenet} as it is made to be challenging with duplicated and dead code samples filtered. First, we extracted a multilingual set composed of code solutions to 237 overlapping \emph{problem\_id}s from C++1400, Python800 and Java250 and it is split into train, validation and test sets following 80\%, 10\% and 10\% sample distributions for each \emph{problem\_id}. For convenience, we name them CodeNetBench-Train/Validation/Test. Furthermore, for the code similarity retrieval task, we test on \emph{CodeNetBench-Sim} set, in which we randomly sample 100 problems and each problem with 10 code snippets in C++ and Python respectively,~\ie 20 code snippets per problem, from CodeNetBench-Test. Finally, we create One-versus-All test pairs,~\ie each test sample is paired with all other samples and positive pairs are those of the same \emph{problem\_id}.}

\noindent\textbf{Tasks. }
\rohan{To evaluate the efficacy of our proposed framework, we evaluate on two tasks as follows:
\begin{itemize}
    \item Code Classification: the goal is to classify source code samples based on their respective programming problem \ruibo{i.e. \emph{problem\_id}}. Code samples belonging to the same programming problem would have high structural and semantic information overlap. As a result, it provides a solid ground for comparing the effectiveness of different source code representations.
    \item Code Similarity: the goal is to compare the efficacy of the various latent sourcecode representations for retrieving \emph{similar} code samples. 
\end{itemize}}
\ruibo{Table \ref{tab:classification_dataset} and \ref{tab:similarity_dataset} summarise the datasets classification model training, evaluation and similarity task evaluation. }~\\
\noindent\textbf{Loss function. } As for the loss function for \emph{problem\_id}s classification, we adopt Additive Angular Margin (AAM) Softmax loss \cite{deng2019arcface}, as shown in Equation \ref{eq:aam}, which has been shown to perform well by explicitly optimising similarity for intra-class samples and diversity for inter-class samples.

\begin{align}
L = -\frac{1}{N}\sum^{N}_{i=1}\ln\frac{\exp\{s\cdot\cos(\theta_{y_i,i} + m)\}}{\exp\{s\cdot\cos(\theta_{y_i,i} + m)\} + \sum_{j\ne y_{i}}\exp\{s \cdot \cos(\theta_{j, i})\}})
\label{eq:aam}
\end{align}
where $\theta_{y_{i}, j}$ is the angle between $i$-th sample to the $y_i$-th class and N is the batch size. We use angular margin penalty $m=0.2$ and feature scale $s=30$. For consistency and fairness, the same loss function is used in all models trained on \emph{problem\_id} classification.

\subsection{Baseline Methods}
\ruibo{
In this section we describe baseline methods to which we compare the proposed method against, including language agnostic and language specific ones. Specifically, we categorise a method as language-agnostic or language-specific 1) if any language-specific pre- or post-processing, including feature extraction, technique is required; 2) if any language-specific assumption is imposed on the model. }

\subsubsection{Language Agnostic Models}~\\

\noindent\textbf{Bag of Characters.} (BoC) \ruibo{A code sample is represented by the relative frequencies of character occurrence. Specifically, the feature vector of a code snippet is formed by counting the number of occurrences of each valid character introduced in \ref{ssec:sourcecode_representation} (excluding \emph{[blank]}). Table \ref{tab:MLP_model_configuration} summarises the configuration of the MLP network that was selected after experiments with an array of setups. The training strategy described in Section \ref{sec:implementation_details} is used.}~\\
\begin{table}[t]
    \centering
    \begin{tabular}{c|c|c|c|c}
    \hline
        \textbf{Input Size} & ~~\textbf{fc0}~~ & ~~\textbf{fc1}~~ & ~~\textbf{fc2}~~ & \textbf{Output} \\
    \hline
    \hline
        $N$ & 128 & 256 & 512 & 237\\
    \hline
    \end{tabular}
    \caption{Bag of Character (BoC) and SPTR-Java MLP model configurations. ReLU activation and BatchNorm are used in fc layers. $N=95$, $256$ and $512$ respectively in BoC, SPTR-Java-S and SPTR-Java-L.}
    \label{tab:MLP_model_configuration}
    
    \begin{tabular}{c|c|c|c|c|c|c}
    \hline
        \textbf{Model} & \textbf{Vocab} & \textbf{Depth} & \textbf{Hidden Size D} & \textbf{MLP size} & \textbf{Heads} & \textbf{Params} \\
    \hline
    \hline
        a-Transformer & 30K & 12 & 128 & 512 & 4 & 7.1M\\
    \hline
        k-Transformer & 120 & 12 & 128 & 512 & 4 & 3.3M\\
    \hline
    \end{tabular}
    \caption{Token transformer model configurations. Learnable position embedding is used.}
    \label{tab:token_transformer_configuration}
\end{table}

\noindent\textbf{Token Transformer.} \ruibo{Similar to CodeBERT \cite{feng-etal-2020-codebert}, relying on the assumption of naturalness in programming language, we build state-of-the-art NLP Transformer with text-based tokens input as baseline. Specifically, we learn two Token Transformer models via supervised training on the \emph{problem\_id} task, one with input of all tokens, including \emph{all} operators and words and another with only 120 combined \emph{keywords} and key operators from Python, Java and C++ provided in \cite{puri2021codenet}, which we call a-Transformer and k-Transformer respectively. To avoid noisy and sparse input embedding for a-Transformer, we extract a vocabulary in which each token occurs at least twice in the training dataset. In addition, both models have maximum input token length of 512. Table \ref{tab:token_transformer_configuration} summarises the configurations of the Transformers that was empirically selected. While the same training strategy described in \ref{sec:implementation_details} is followed, due to the higher model memory footprint compared to other models, we use a smaller batch size of 32.}

\subsubsection{Language Specific Models}~\\

\noindent\textbf{Simplified Parse Tree Relation.} (SPTR) \ruibo{Leveraging Simplified Parse Tree (SPT) features originally presented in \cite{luan2019}, SPTR directly exploits the underlying structure from code snippets. We extract SPT and build a vocabulary from all training Java code samples. Then a sparse binary count vector is extracted from each SPT that defines the existence of a particular vocabulary from a SPT. Finally dense feature vectors are obtained through Truncated Single Value Decomposition (SVD). We experiment with the two MLP models,~\ie SPTR-Java-S and SPTR-Java-L, with Truncated SVD output dimensions set to 256 and 512 respectively with explained variance ratios of 0.768 and 0.834. Model configuration is summarised in Table \ref{tab:MLP_model_configuration}. We train and report results on all Java samples in CodeNetBench-Train and CodeNetBench-Test, following the same training strategy described in section \ref{sec:implementation_details}.}

\subsection{CV4Code setup}
\renewcommand{\arraystretch}{1.05}
\begin{table}[t]
    \centering
    \begin{tabular}{c|c|c|c|c|c}
    \hline
        \textbf{conv0} & \textbf{conv1} & \textbf{conv2} & \textbf{conv3} & \textbf{fc} & \textbf{Output} \\
    \hline
    \hline
        $16, 2$ & $\begin{bmatrix} 64\\ 64 \end{bmatrix} \times 2$, 2 & $\begin{bmatrix} 128\\128 \end{bmatrix} \times 2$, 2 & $\begin{bmatrix} 256\\256\end{bmatrix} \times 2, 1$ & $128$ & $237$\\
    \hline
    \end{tabular}
    \caption{CV4Code-ResNet model configuration. Conv layer weights are annotated as number of filters and stride step size. $7\times7$ kernel is used in conv0 and $3\times3$ in others. $3\times 3$ with stride 2 and $6\times6$ (global) max\_pool is used after conv0 and conv3. Total \#params=3.25M.}
    \label{tab:resnet_configuration}
\end{table}
\begin{table}[t]
    \centering
    \begin{tabular}{c|c|c}
    \hline
        \textbf{Model} & \textbf{Patch size} & \textbf{Params} \\
    \hline
    \hline
        CV4Code-ViT-S & $16\times16$ & 5.32M\\
    \hline
        CV4Code-ViT-L & $8\times8$ & 2.98M\\
    \hline
        CV4Code-ViT-fsd-S & $16\times16$ & 13.97M\\
    \hline
        CV4Code-ViT-fsd-L & $8\times8$ & 4.58M\\
    \hline
    \end{tabular}
    \caption{CV4Code-ViT and CV4Code-ViT-fsd configurations. All configurations use depth of 8, hidden size of 128, MLP size of 512 with 4 heads. Learnable position embedding is used. }
    \label{tab:token_transformer_configuration}
    
    \renewcommand{\arraystretch}{1.05}
    \begin{tabular}{c|c|c|c}
    \hline
        \textbf{Model} & \textbf{Convolutional Tokenizer} & \textbf{Visual Tokens} & \textbf{Params} \\
    \hline
    \hline
        CV4Code-CCT-S & $\begin{bmatrix} 7\times 7, 64\end{bmatrix} \times 2$, stride 2 & 49 & 2.35M\\
    \hline
        CV4Code-CCT-L & $\begin{bmatrix} 3\times 3, 64\end{bmatrix} \times 3$, stride 1 & 169 & 5.3M\\
    \hline
    \end{tabular}
    \caption{CV4Code-CCT configurations. All configurations use depth of 8, hidden size of 128, MLP size of 512 with 4 heads, no position embedding. Each convolutional layer is followed by a $2\times2$ max\_pool layer with stride of 2. Fixed Sinusoidal position embedding is used.}
    \label{tab:vit_configuration}
\end{table}
Table \ref{tab:resnet_configuration} reports the model configuration for CV4Code-ResNet, we experiment with a number of configurations and report result in this paper from the model with the best validation accuracy. For CV4Code-ViT and ViT-fsd, we experiment with two different patch sizes,~\ie $16\times16$ and $8\times8$, which result in 36 and 144 visual tokens respectively, given input sourcecode image of size $96\times96$. For CV4Code-CCT, we compare two convolutional soft tokenization setups which result in similar length of visual tokens. Table \ref{tab:vit_configuration} summarises the configurations for all vision transformer variants.

\subsection{Results}
\begin{table}[t]
    \centering
    \begin{tabular}{c|c|c||c|c}
    \hline
        \multirow{2}{*}{\textbf{Model}} & \multicolumn{2}{c||}{\textbf{Multilingual}} & \multicolumn{2}{c}{\textbf{Java-only}}\\
         & ~~~Top-1~~~ & ~~~Top-5~~~  & ~~~Top-1~~~ & ~~~Top-5~~~ \\
        \hline\hline
        BoC & 80.97 & 90.16 & 80.56 & 89.74 \\
        k-Transformer & 90.30 & 95.42 & 89.54 & 94.97 \\
        a-Transformer & 93.58 & 96.63 & 94.04 & 97.40 \\
        SPTR-Java-S & - & - & 91.09 & 96.98 \\
        SPTR-Java-L & - & - & 92.95 & 96.78 \\
        \hdashline
        ResNet & 92.93 & 96.50 & 91.17 & 95.50\\
        ViT-S & 85.45 & 93.64 & 80.50 & 90.95 \\
        ViT-L & 92.85 & 96.86 & 90.27 & 95.46 \\
        ViT-fsd-S & 86.04 & 93.80 & 80.60 & 90.80 \\
        ViT-fsd-L & 92.27 & 96.47 & 88.99 & 94.49\\
        CCT-S & 96.08 & 98.45 & 94.63 & 98.01 \\
        CCT-L & \textbf{97.64} & \textbf{98.99} & \textbf{97.13} & \textbf{98.79} \\
    \hline
    \end{tabular}
    \caption{\emph{problem\_id} classification results on CodeNetBench-Test.}
    \label{tab:classification_results}
\end{table}
\begin{table}[t]
    \centering
    \begin{tabular}{c|c}
    \hline
        \textbf{Model} & \textbf{mAP@R}\\
        \hline\hline
        a-Transformer & 0.980 \\
        ResNet & 0.983 \\
        CCT-L & \textbf{0.991} \\
    \hline
    \end{tabular}
    \caption{Code similarity evaluation result on CodeNetBench-Sim.}
    \label{tab:code_similarity_results}
\end{table}
\noindent\textbf{Evaluation Metrics. }
\ruibo{For classification tasks, top-1 and top-5 accuracy are reported as the evaluation metrics on CodeNetBench-Test. For code similarity, we consider a retrieval task where a code snippet is used an query to search for similar snippets from a pool and mAP@R\cite{musgrave2020mapatr} is reported as the main evaluation metric. }
\begin{figure}[t]
    \centering
    \includegraphics[width=\textwidth]{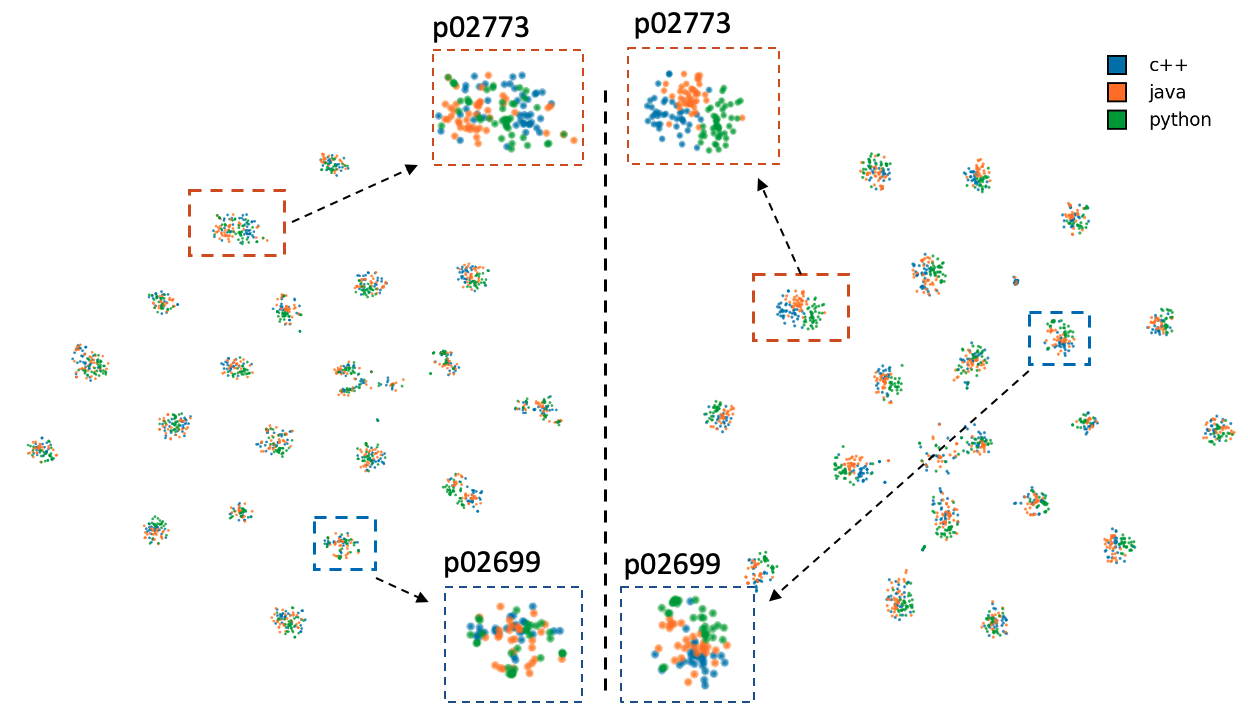}
    \caption{t-SNE 2D projection of \textbf{left}: CV4Code-CCT-L and \textbf{right}: a-Transformer embeddings. Colour-labeled by unique programming \emph{language}s.}
    \label{fig:tsne_language_1}
\end{figure}

\subsubsection{Language-agnostic Classification.}
\ruibo{As summarised in Table \ref{tab:classification_results}, it can be seen from multilingual test result that CV4Code-CCT-L outperforms all other models in terms of both top-1 and top-5 accuracy, including the strong NLP-based baseline a-Transformer. We attribute the gain of CV4Code over k-Transformer to its ability to exploit the contextual and structural information embedded in the spatial relationships, which is not directly available through the sequential token input in k-Transformer. Comparing CV4Code ResNet and transformer variants, ViT and ViT-fsd variants all perform worse than ResNet. This potentially implies that the inductive bias,~\eg locality, associated with convolutional networks are critical for sourcecode understanding from our proposed sourcecode image representation. This is further supported by the gain obtained in CCT-S/L which inherit the inductive bias through its soft convolutional tokenizer and leverage the expressiveness of the Transformer network. In addition, it is noticed that the size of patch and the subsequently generated visual tokens have a strong influence on the performance of all CV4Code transformer variants, which is exhibited through the gap between all -S/-L model pairs.}

\subsubsection{Language-specific (Java) Classification}\label{sec:language_specific_results}
\ruibo{The Java-only column in Table \ref{tab:classification_results} summaries model test results on Java samples in CodeNetBench-Test. Although SPTR-Java-L/S both achieve strong performance, which demonstrates the effectiveness of the input features that exploit the underlying code structures via SPTs, it is outperformed by a-Transformer by $1.09\%$ in terms of Top-1 accuracy. And CV4Code-CCT-L, as a language-agnostic model trained on C++, Python and Java, still achieves the strongest result overall with a $97.13\%$ Top-1 accuracy.}

\subsubsection{Code Similarity}
\ruibo{We test the learned embeddings from the models that achieve strong results on the classification task,  with respect to code similarity and report similarity mAP@R scores on datasets summarised in Table \ref{tab:similarity_dataset}. We extract \emph{[class]} embedding at the transformer output from a-Transformer,  pre-ReLU bottleneck output from CV4Code-ResNet and sequence pooled embedding from transformer output from CV4Code-CCT-L. Cosine similarity is used to compute the pairwise similarity of the embeddings. Test results are shown in Table \ref{tab:code_similarity_results}. We observe that CV4Code models, including ResNet and CCT-L, outperform a-Transformer on this task. This implies that the learned embeddings from the proposed CV4Code models are highly discriminative and encodes the \emph{semantics} of sourcecode, which in this case is the problem that a code snippet is trying to solve. 
}

\subsection{Ablation studies}

\noindent\textbf{Influence of Programming Languages}. 
\ruibo{We look at the influence on programming languages on the generated latent sourcecode representations by CV4Code-CCT-L model, in comparison with a-Transformer. For visualisation, we use t-SNE\cite{laurens2008tsne} to project the learned embeddings and colour code the data points by its programming language. As shown in figure \ref{fig:tsne_language_1}, while in both embedding spaces clusters are formed with respect to \emph{problem\_id}s following the training task, we observe slightly more obvious sub-clusters with respect to \emph{languages} in a-Transformer, especially Python against Java and C++. This potentially suggests that token sequence model such as a-Transformer, with text-based token input sequence, is more sensitive to the underlying programming language compared to the proposed CV4Code approach. We suspect that this is due to the unique syntax, use of operators, specific coding styles and naming conventions often adopted in each programming language,~\eg different variable naming conventions, which has more influence on the overall snippet-level representation of a-Transformer in which token-level embeddings are used. }

\noindent\textbf{Future Studies}.
\ruibo{Following recent advancement of self-supervised learning for Vision Transformers \cite{he2021masked}, we would like to test and scale up the CV4Code transformer with the abundance of unlabelled sourcecode snippets in the public domain. With such a setup, we look to establish a study of self-supervised approaches with the proposed CV4Code method and compare to NLP approaches,\eg BERT, for additional MLOnCode problems. Furthermore, to capture richer semantics in the latent space and test on more generic downstream applications,~\eg code quality assessment, we plan to extend the current training framework to include a multi-task learning configuration, from scratch or finetuned from a pretrained model. }
\section{Conclusion}
\label{sec:conclusion}
\ruibo{In this work, we propose an idea for sourcecode understanding via a novel visual representation. The proposed method treats sourcecode as images and uses supervised training to predict the programming problem that the sourcecode snippet is solving. Compared to traditional syntax parse tree or token based methods, the proposed approach is programming language agnostic and does not depend on syntax correctness in the preprocessing stage. Finally, along with a thorough study of vision models applied on the proposed representations and a comparison with syntax tree and NLP-based approaches, using a Compact Convolution Transformer CV4Code model, we show state-of-the-art performance in terms of both classification and code similarity retrieval. }
%
%
\bibliographystyle{splncs04}
\bibliography{egbib}

\begin{thebibliography}{10}
\providecommand{\url}[1]{\texttt{#1}}
\providecommand{\urlprefix}{URL }
\providecommand{\doi}[1]{https://doi.org/#1}

\bibitem{Allamanis13}
{Allamanis}, M., {Sutton}, C.: Mining source code repositories at massive scale
  using language modeling. In: 2013 10th Working Conference on Mining Software
  Repositories (MSR). pp. 207--216 (2013). \doi{10.1109/MSR.2013.6624029}

\bibitem{allamanis2018survey}
Allamanis, M., Barr, E.T., Devanbu, P., Sutton, C.: A survey of machine
  learning for big code and naturalness. ACM Computing Surveys (CSUR)
  \textbf{51}(4), ~81 (2018)

\bibitem{alon2018codeseq}
Alon, U., Brody, S., Levy, O., Yahav, E.: code2seq: Generating sequences from
  structured representations of code. In: International Conference on Learning
  Representations (2019), \url{https://openreview.net/forum?id=H1gKYo09tX}

\bibitem{Uri19}
Alon, U., Zilberstein, M., Levy, O., Yahav, E.: Code2vec: Learning distributed
  representations of code. Proc. ACM Program. Lang.  \textbf{3}(POPL) (Jan
  2019). \doi{10.1145/3290353}, \url{https://doi.org/10.1145/3290353}

\bibitem{bilgin2021code2image}
Bilgin, Z.: Code2image: Intelligent code analysis by computer vision techniques
  and application to vulnerability prediction. arXiv preprint arXiv:2105.03131
  (2021)

\bibitem{Chen21}
Chen, Z., Kommrusch, S., Tufano, M., Pouchet, L.N., Poshyvanyk, D., Monperrus,
  M.: Sequencer: Sequence-to-sequence learning for end-to-end program repair.
  IEEE Transactions on Software Engineering  \textbf{47}(9),  1943--1959
  (2021). \doi{10.1109/TSE.2019.2940179}

\bibitem{deng2019arcface}
Deng, J., Guo, J., Xue, N., Zafeiriou, S.: Arcface: Additive angular margin
  loss for deep face recognition. In: Proceedings of the IEEE/CVF conference on
  computer vision and pattern recognition. pp. 4690--4699 (2019)

\bibitem{dey2019socodecnn}
Dey, S., Singh, A.K., Prasad, D.K., Mcdonald-Maier, K.D.: Socodecnn: Program
  source code for visual cnn classification using computer vision methodology.
  IEEE Access  \textbf{7},  157158--157172 (2019)

\bibitem{feng-etal-2020-codebert}
Feng, Z., Guo, D., Tang, D., Duan, N., Feng, X., Gong, M., Shou, L., Qin, B.,
  Liu, T., Jiang, D., Zhou, M.: {C}ode{BERT}: A pre-trained model for
  programming and natural languages. In: Findings of the Association for
  Computational Linguistics: EMNLP 2020. pp. 1536--1547. Association for
  Computational Linguistics, Online (Nov 2020).
  \doi{10.18653/v1/2020.findings-emnlp.139},
  \url{https://aclanthology.org/2020.findings-emnlp.139}

\bibitem{Hassani2021EscapingTB}
Hassani, A., Walton, S., Shah, N., Abuduweili, A., Li, J., Shi, H.: Escaping
  the big data paradigm with compact transformers. ArXiv
  \textbf{abs/2104.05704} (2021)

\bibitem{he2021masked}
He, K., Chen, X., Xie, S., Li, Y., Doll{\'a}r, P., Girshick, R.: Masked
  autoencoders are scalable vision learners. arXiv preprint arXiv:2111.06377
  (2021)

\bibitem{he2016residual}
He, K., Zhang, X., Ren, S., Sun, J.: Deep residual learning for image
  recognition. In: 2016 IEEE Conference on Computer Vision and Pattern
  Recognition (CVPR). pp. 770--778 (2016). \doi{10.1109/CVPR.2016.90}

\bibitem{alex2021vit}
Kolesnikov, A., Dosovitskiy, A., Weissenborn, D., Heigold, G., Uszkoreit, J.,
  Beyer, L., Minderer, M., Dehghani, M., Houlsby, N., Gelly, S., Unterthiner,
  T., Zhai, X.: An image is worth 16x16 words: Transformers for image
  recognition at scale (2021)

\bibitem{Lee2021VisionTF}
Lee, S.H., Lee, S., Song, B.C.: Vision transformer for small-size datasets.
  arXiv preprint abs/2112.13492  (2021)

\bibitem{Loshchilov2019DecoupledWD}
Loshchilov, I., Hutter, F.: Decoupled weight decay regularization. In: ICLR
  (2019)

\bibitem{luan2019}
Luan, S., Yang, D., Barnaby, C., Sen, K., Chandra, S.: Aroma: Code
  recommendation via structural code search. Proc. ACM Program. Lang.
  \textbf{3}(OOPSLA) (Oct 2019)

\bibitem{laurens2008tsne}
van~der Maaten, L., Hinton, G.: Visualizing data using t-sne. Journal of
  Machine Learning Research  \textbf{9}(86),  2579--2605 (2008),
  \url{http://jmlr.org/papers/v9/vandermaaten08a.html}

\bibitem{musgrave2020mapatr}
Musgrave, K., Belongie, S., Lim, S.N.: A metric learning reality check. In:
  Vedaldi, A., Bischof, H., Brox, T., Frahm, J.M. (eds.) Computer Vision --
  ECCV 2020. pp. 681--699. Springer International Publishing, Cham (2020)

\bibitem{ochodek2020recognizing}
Ochodek, M., Hebig, R., Meding, W., Frost, G., Staron, M.: Recognizing lines of
  code violating company-specific coding guidelines using machine learning.
  Empirical Software Engineering  \textbf{25}(1),  220--265 (2020)

\bibitem{Pradel2018}
Pradel, M., Sen, K.: Deepbugs: A learning approach to name-based bug detection.
  Proc. ACM Program. Lang.  \textbf{2}(OOPSLA) (Oct 2018).
  \doi{10.1145/3276517}, \url{https://doi.org/10.1145/3276517}

\bibitem{puri2021codenet}
Puri, R., Kung, D., Janssen, G., Zhang, W., Domeniconi, G., Zolotov, V., Dolby,
  J., Chen, J., Choudhury, M., Decker, L., Thost, V., Buratti, L., Pujar, S.,
  Finkler, U.: Project codenet: A large-scale ai for code dataset for learning
  a diversity of coding tasks (2021)

\bibitem{rabin2021encoding}
Rabin, M.R.I., Alipour, M.A.: Encoding program as image: Evaluating visual
  representation of source code. arXiv preprint arXiv:2111.01097  (2021)

\bibitem{Russell18}
Russell, R.L., Kim, L.Y., Hamilton, L.H., Lazovich, T., Harer, J.A., Ozdemir,
  O., Ellingwood, P.M., McConley, M.W.: Automated vulnerability detection in
  source code using deep representation learning. CoRR  \textbf{abs/1807.04320}
  (2018), \url{http://arxiv.org/abs/1807.04320}

\bibitem{sharma2021survey}
Sharma, T., Kechagia, M., Georgiou, S., Tiwari, R., Sarro, F.: A survey on
  machine learning techniques for source code analysis. arXiv preprint
  arXiv:2110.09610  (2021)

\bibitem{Simonyan15}
Simonyan, K., Zisserman, A.: Very deep convolutional networks for large-scale
  image recognition. In: International Conference on Learning Representations
  (2015)

\bibitem{Svyatkovskiy21}
Svyatkovskiy, A., Lee, S., Hadjitofi, A., Riechert, M., Franco, J.V.,
  Allamanis, M.: Fast and memory-efficient neural code completion. In: 18th
  {IEEE/ACM} International Conference on Mining Software Repositories, {MSR}
  2021, Madrid, Spain, May 17-19, 2021. pp. 329--340. {IEEE} (2021).
  \doi{10.1109/MSR52588.2021.00045},
  \url{https://doi.org/10.1109/MSR52588.2021.00045}

\bibitem{Shuhan20}
Yan, S., Yu, H., Chen, Y., Shen, B., Jiang, L.: Are the code snippets what we
  are searching for? a benchmark and an empirical study on code search with
  natural-language queries. In: 2020 IEEE 27th International Conference on
  Software Analysis, Evolution and Reengineering (SANER). pp. 344--354 (2020).
  \doi{10.1109/SANER48275.2020.9054840}

\bibitem{Zanoni15}
Zanoni, M., Arcelli~Fontana, F., Stella, F.: On applying machine learning
  techniques for design pattern detection. J. Syst. Softw.  \textbf{103}(C),
  102–117 (May 2015). \doi{10.1016/j.jss.2015.01.037},
  \url{https://doi.org/10.1016/j.jss.2015.01.037}

\bibitem{zhang2019novel}
Zhang, J., Wang, X., Zhang, H., Sun, H., Wang, K., Liu, X.: A novel neural
  source code representation based on abstract syntax tree. In: 2019 IEEE/ACM
  41st International Conference on Software Engineering (ICSE). pp. 783--794.
  IEEE (2019)

\bibitem{zuegner_code_transformer_2021}
Z{\"u}gner, D., Kirschstein, T., Catasta, M., Leskovec, J., G{\"u}nnemann, S.:
  Language-agnostic representation learning of source code from structure and
  context. In: International Conference on Learning Representations (ICLR)
  (2021)

\bibitem{zugner2021language}
Z{\"u}gner, D., Kirschstein, T., Catasta, M., Leskovec, J., G{\"u}nnemann, S.:
  Language-agnostic representation learning of source code from structure and
  context. arXiv preprint arXiv:2103.11318  (2021)

\end{thebibliography}
\end{document}